\documentclass[conference]{IEEEtran}
\IEEEoverridecommandlockouts
\usepackage{cite}
\usepackage{amsmath,amssymb,amsfonts}
\usepackage{amsmath,epsfig}
\usepackage{algorithm,algorithmic}
\usepackage{multirow}
\usepackage{booktabs}
\usepackage{graphicx}
\usepackage{float}
\usepackage{caption}
\usepackage{textcomp}
\usepackage{xcolor}
\usepackage{titlesec}

\let\OLDthebibliography\thebibliography
\renewcommand\thebibliography[1]{
  \OLDthebibliography{#1}
  \setlength{\parskip}{0pt}
  \setlength{\itemsep}{0pt plus 0.3ex}
}

\def\x{{\mathbf x}}
\def\L{{\cal L}}

\def\BibTeX{{\rm B\kern-.05em{\sc i\kern-.025em b}\kern-.08em
    T\kern-.1667em\lower.7ex\hbox{E}\kern-.125emX}}
\begin{document}

\title{Time-Frequency Jointed Imperceptible Adversarial Attack to Brainprint Recognition with Deep Learning Models\\
\thanks{This work was supported by National Natural Science Foundation of China (U20B2074), Key Research and Development Project of Zhejiang Province (2023C03026, 2021C03001, 2021C03003), and supported by Key Laboratory of Brain Machine Collaborative Intelligence of Zhejiang Province (2020E10010).}
}

\author{\IEEEauthorblockN{Hangjie Yi\textsuperscript{1,2}, Yuhang Ming\textsuperscript{1,2}, Dongjun Liu\textsuperscript{1,2}, Wanzeng Kong\textsuperscript{*1,2}\thanks{\textsuperscript{*} Wanzeng Kong is the corresponding author.}}

\IEEEauthorblockA{\textsuperscript{1}\textit{School of Computer Science}, \textit{Hangzhou Dianzi University}, Hangzhou, Zhejiang, China}
\IEEEauthorblockA{\textsuperscript{2}\textit{Key Laboratory of Brain Machine Collaborative Intelligence of Zhejiang Province}, Hangzhou, Zhejiang, China}
\IEEEauthorblockA{\{yihangjie, yuhang.ming, liudongjun, kongwanzeng\}@hdu.edu.cn}}

\maketitle

\begin{abstract}
EEG-based brainprint recognition with deep learning models has garnered much attention in biometric identification. Yet, studies have indicated vulnerability to adversarial attacks in deep learning models with EEG inputs. In this paper, we introduce a novel adversarial attack method that jointly attacks time-domain and frequency-domain EEG signals by employing wavelet transform. Different from most existing methods which only target time-domain EEG signals, our method not only takes advantage of the time-domain attack's potent adversarial strength but also benefits from the imperceptibility inherent in frequency-domain attack, achieving a better balance between attack performance and imperceptibility. Extensive experiments are conducted in both white- and grey-box scenarios and the results demonstrate that our attack method achieves state-of-the-art attack performance on three datasets and three deep-learning models. In the meanwhile, the perturbations in the signals attacked by our method are barely perceptible to the human visual system.
\end{abstract}

\begin{IEEEkeywords}
EEG, brainprint, adversarial attack, biometrics, wavelets transform
\end{IEEEkeywords}

\section{Introduction}
\label{sec:intro}
Brainprint recognition based on electroencephalogram (EEG) signals has gradually attracted increasing attention as a biometric identification technique~\cite{zhangReviewEEGBasedAuthentication2021}. Various convolution-based deep neural networks have been employed to capture EEG features that magnify inter-subject variations and minimize intra-subject discrepancies, thereby achieving superior cross-task and cross-session identity recognition performance~\cite{maoEEGbasedBiometricIdentification2017}~\cite{chenEEGbasedBiometricIdentification2020}~\cite{wilaiprasitpornAffectiveEEGBasedPerson2020}~\cite{jinCTNNConvolutionalTensorTrain2021}~\cite{kumarEvidenceTaskIndependentPersonSpecific2021}. Such models can also mitigate the influence of task-related information in EEG features for identity recognition. Deep learning models have achieved state-of-the-art performance in brainprint recognition, eliminating the need for manual feature extraction and facilitating end-to-end brainprint recognition. Notably, the EEGNet~\cite{lawhernEEGNetCompactConvolutional2018} proposed in 2018, as well as the DeepConvNet and ShallowConvNet~\cite{schirrmeisterDeepLearningConvolutional2017} introduced in 2017, stand out as exemplary models. They have tailored convolutional kernels to extract spatiotemporal EEG features, and are extensively used as backbone architectures in various models.

\begin{figure}[t]
  \centering
  \includegraphics[width=\linewidth]{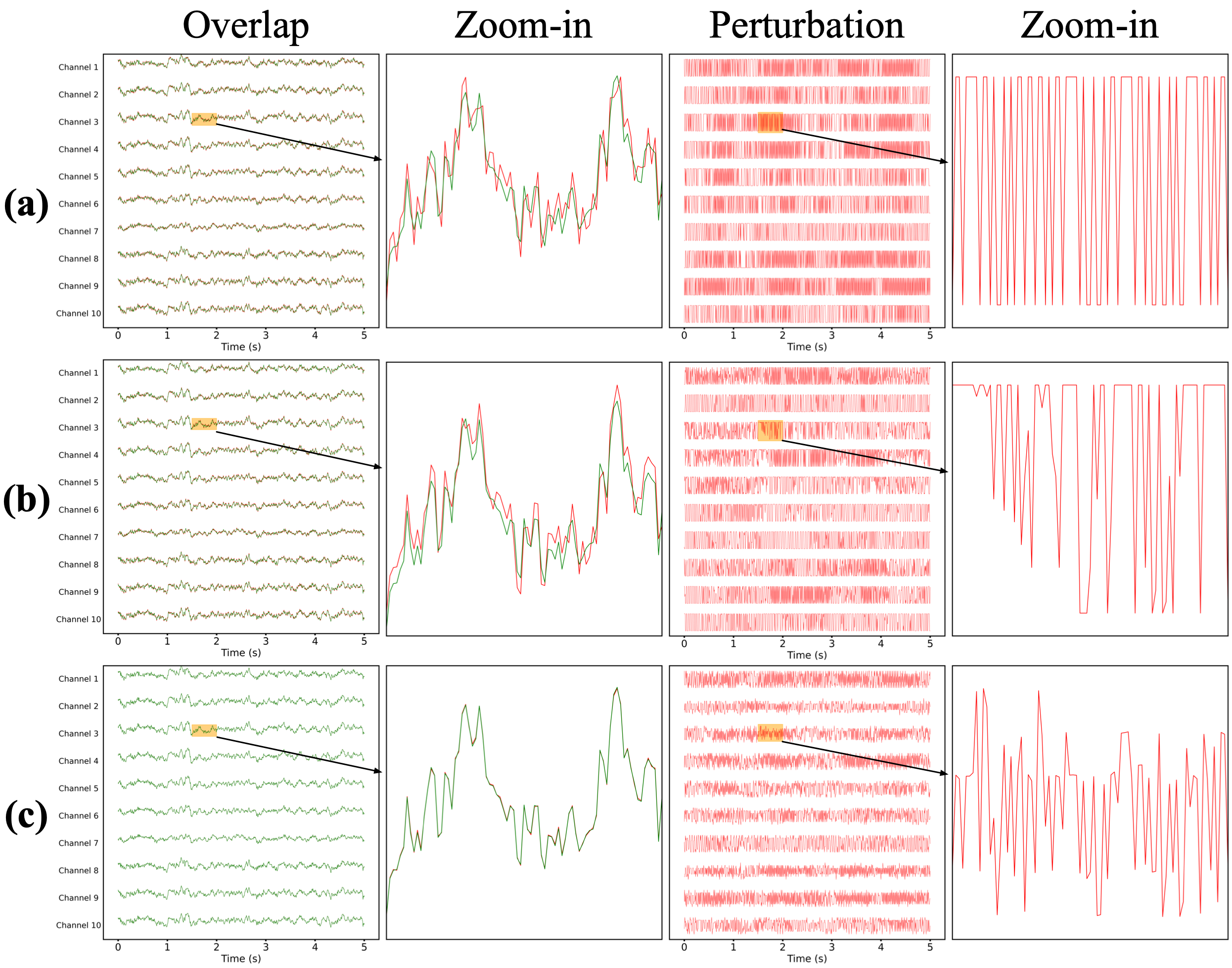}
  \captionsetup{font=small}
  \caption{The comparison between benign and adversarial brainprint examples attacked by (a)FGSM, (b)PGD, and (c)our TFAttack.
  From left to right we present the overlap of the benign examples (green) and the corresponding adversarial examples (red), zoom-in of the overlapped examples, the perturbations added in the benign examples, and the zoom-in of the perturbations.
  Because of the square-wave patterns, the difference between the adversarial examples (red) attacked by FGSM and PGD and the benign examples (green) are much more evident compared to the ones from our proposed TFAttack.
  }
\label{fig:comparision}
\vspace{-2em}
\end{figure}

However, recent research indicates that EEG-based brain-computer interfaces (BCIs) exhibit vulnerabilities to adversarial attacks~\cite{zhangVulnerabilityCNNClassifiers2019}.
Deep learning models used for BCIs can be deceived by adversarial examples, resulting in a significant drop in model's performance and even causing harm or fatigue to users~\cite{zhangVulnerabilityCNNClassifiers2019}~\cite{fengSagaSparseAdversarial2021}~\cite{bianSSVEPbasedBraincomputerInterfaces2022}~\cite{wangPhysicallyConstrainedAdversarialAttacks2022}. 
Through hijacking EEG signals, attackers can disrupt the operation of steady-state visually evoked potentials (SSVEP) based typewriters~\cite{bianSSVEPbasedBraincomputerInterfaces2022} or control wheelchairs~\cite{liMultimodalBCIsTarget2016}, leading to unintended characters or dangerous movements, and endangering user safety. Adversarial attacks could also hinder the application of brainprint recognition in security-sensitive scenarios. Attackers could block legitimate users from accessing the system and illegally modify critical data, causing harm to organizations or individuals. To the best of our knowledge, there has been no research on adversarial attacks for brainprint recognition, emphasizing the urgency and significance of such studies.

In 2019, Zhang \textit{et al.} discovered the fragility of CNN networks used in MI classification, P300 detection, and ERN classification, introducing the unsupervised fast gradient sign method (UFGSM) attack algorithm~\cite{zhangVulnerabilityCNNClassifiers2019}. Existing adversarial attacks specific to EEG-based BCIs models, such as fast gradient sign method (FGSM)~\cite{goodfellowExplainingHarnessingAdversarial2015a} and projected gradient descent (PGD)~\cite{madryDeepLearningModels2018}, exploit gradient information of input data while also using constraints like the infinity norm to generate tiny adversarial perturbations. However, the perturbations generated by these methods in the time domain resemble square waves, having abrupt ascending and descending edges, which are perceptible to the human visual system (HVS) as demonstrated in Fig.~\ref{fig:comparision}. Additionally, such attack methods overlook the crucial frequency information in EEG signals, which significantly influences model decision-making in paradigms without prominent features like brainprint recognition and emotion classification.

To address the aforementioned challenges, we propose an imperceptible adversarial attack jointly based on time-frequency domains (TFAttack) for brainprint recognition. Our approach employs wavelet transform to satisfy the transitions between the time and frequency domains of EEG signals and is compatible with backpropagation. Initially, we convert the input EEG time-domain signal to the frequency domain using the wavelet transform, followed by an inverse wavelet transform to revert to the time domain. Subsequently, we utilize the target model to infer from the time-domain signal, compute its adversarial loss, derive the gradient of the frequency-domain signal, and optimize frequency-domain perturbation. To further optimize the efficiency of the adversarial attack, we adopt an alternating strategy, optimizing adversarial perturbations between the time and frequency domains. Besides $L_2$ norm and cosine similarity used to evaluate the imperceptibility of adversarial examples, we also employ the Dynamic Time Warping (DTW) score\cite{cuturiSoftDTWDifferentiableLoss2017} as an imperceptibility metric. The DTW score can evaluate the geometric similarity between two waves, better reflecting HVS's preferences.

\begin{itemize}
    \item To our best knowledge, this work is the first to explore adversarial attacks in brainprint recognition.
  \item Our proposed TFAttack harnesses both time and frequency domain data, producing powerful and imperceptible adversarial examples.
  \item We have conducted experiments across three datasets and three deep learning models, achieving state-of-the-art attack performance in white-box and gray-box scenarios.
\end{itemize}

\section{Related Works}

\subsection{Brainprint Recognition}
With the rapid development of deep learning across various fields in recent years, numerous convolutional neural network (CNN) based models have been applied to EEG decoding. In 2017, ~\cite{schirrmeisterDeepLearningConvolutional2017} introduced DeepConvNet and ShallowConvNet, designing temporal and spatial convolutional kernels that achieved outstanding performance across multiple BCI applications. In 2018, ~\cite{lawhernEEGNetCompactConvolutional2018} proposed EEGNet, a model based on spatiotemporal separable convolution, which is widely recognized in EEG decoding. EEGNet reduced the possibility of overfitting by minimizing model parameters. Increasing efforts have been dedicated to using deep learning models for brainprint recognition in recent years. ~\cite{maoEEGbasedBiometricIdentification2017} presented a CNN that directly validates the identity of 100 subjects on raw EEG signals in 2017. ~\cite{chenEEGbasedBiometricIdentification2020} introduced a CNN model called GSLT-CNN, which utilizes global spatial and local temporal filters. ~\cite{wilaiprasitpornAffectiveEEGBasedPerson2020} presented a model dedicated to EEG-based identity verification using CNN and recurrent neural network (RNN). ~\cite{jinCTNNConvolutionalTensorTrain2021} proposed a neural network based on convolution and tensor training, achieving an accuracy rate exceeding 99\% on cross-task brainprint datasets. ~\cite{cuiEEGAuthenticationBased2022} incorporated attention mechanisms and a triple loss function into the traditional cascade network based on CNN to enhance performance. Essentially, deep learning-based brainprint recognition models predominantly employ CNN-based architectures, with DeepConvNet, ShallowConvNet, and EEGNet frequently serving as the backbone for feature extraction\cite{chenEEGbasedBiometricIdentification2020}\cite{wilaiprasitpornAffectiveEEGBasedPerson2020}\cite{jinCTNNConvolutionalTensorTrain2021}\cite{fallahiBrainNetImprovingBrainwavebased2023b}. Therefore, we selected DeepConvNet, ShallowConvNet, and EEGNet as the target models for adversarial attacks.

\subsection{Adversarial Attacks in BCIs}  
Recent deep-learning models have achieved high accuracy in processing images, audio, language, and EEG signals. However, ~\cite{goodfellowExplainingHarnessingAdversarial2015a} discovered that adding tiny perturbations to image data can lead neural network models to misclassify, and the FGSM attack algorithm was proposed. Subsequently, algorithms based on FGSM, such as PGD~\cite{madryDeepLearningModels2018}, basic iterative method (BIM)~\cite{kurakinAdversarialExamplesPhysical2017a}, and C\&W~\cite{carliniEvaluatingRobustnessNeural2017} were proposed. These algorithms are proposed for attacking image data. In 2019, ~\cite{zhangVulnerabilityCNNClassifiers2019} found that attacking raw EEG signals could significantly degrade models' performance, and proposed the UFGSM attack algorithm which generates perturbation that is similar to square wave. ~\cite{fengSagaSparseAdversarial2021} introduced the Sparse Adversarial eeG Attack (SAGA) based on PGD, which only attacks specific channels and time segments of EEG data. ~\cite{bianSSVEPbasedBraincomputerInterfaces2022} discovered that BCIs based on SSVEP are susceptible to square wave attacks. ~\cite{wangPhysicallyConstrainedAdversarialAttacks2022} also proposed a physically constrained adversarial attack based on PGD, demonstrating the vulnerability of the EEGNet model in motor imagery. In summary, there are currently no adversarial attack algorithms specifically for brainprint recognition in BCIs, and existing algorithms do not consider the frequency information of EEG signals, only relying on time-domain attack algorithms such as PGD.

\section{Method}
In the white-box attack scenario, a malicious attacker knows all the details about the target model, including its architecture, parameters, loss function, gradients, and outputs for given inputs. Leveraging this knowledge, the attacker crafts adversarial examples $x_{adv}$ that are deliberately calculated to deceive the target model and degrade its classification performance. Distance metrics such as $L_2$ and $L_{\infty}$ are commonly utilized to quantify the dissimilarity between benign samples and their adversarial counterparts. The scenario of white-box attacks can be described as:
\begin{equation}\label{eq1}
    \arg \max D\left(x^{\mathrm{adv}}, x^{\mathrm{benign}}\right) \text {, s.t. } f(x^{\mathrm{adv}}) \neq y^{\mathrm{gt}} \text {, }
\end{equation}
where $D$ is a distance metric function characterizing the similarity between the benign example and adversarial example, $f$ denotes the target model and $y^{\mathrm{gt}}$ denotes the ground truth.

In this work, we introduce an attack method for brainprint recognition that operates in both time and frequency domains (TFAttack). This approach is a composite of time-domain attacks (TAttack) and frequency-domain attacks (FAttack), capitalizing on gradients from the time-domain and frequency-domain data to achieve a more imperceptible and powerful attack. The TAttack directly generates perturbation on raw EEG signals, while the FAttack focuses on attacking the four frequency components derived from wavelet transformation.

\subsection{Frequency Domain Attack}

Distinct from SSVEP or event-related potential (ERP) which exhibit prominent signal characteristics, such as periodic brain responses and P300 signals, individuals' pattern features of brainprint are elusive to discern. They potentially embody a time-frequency hybrid characteristic, lending some rationale to the feasibility of attacking brainprint signals in the frequency domain. Furthermore, perturbations generated through time-domain attacks typically manifest as square wave signals, making them more perceptible to the HVS. In contrast, perturbations resulting from frequency-domain attacks, when observed in the time domain, appear more like random signals, leading to a more invisible attack.

\begin{figure}[t]
  \centering
  \includegraphics[width=0.9\linewidth]{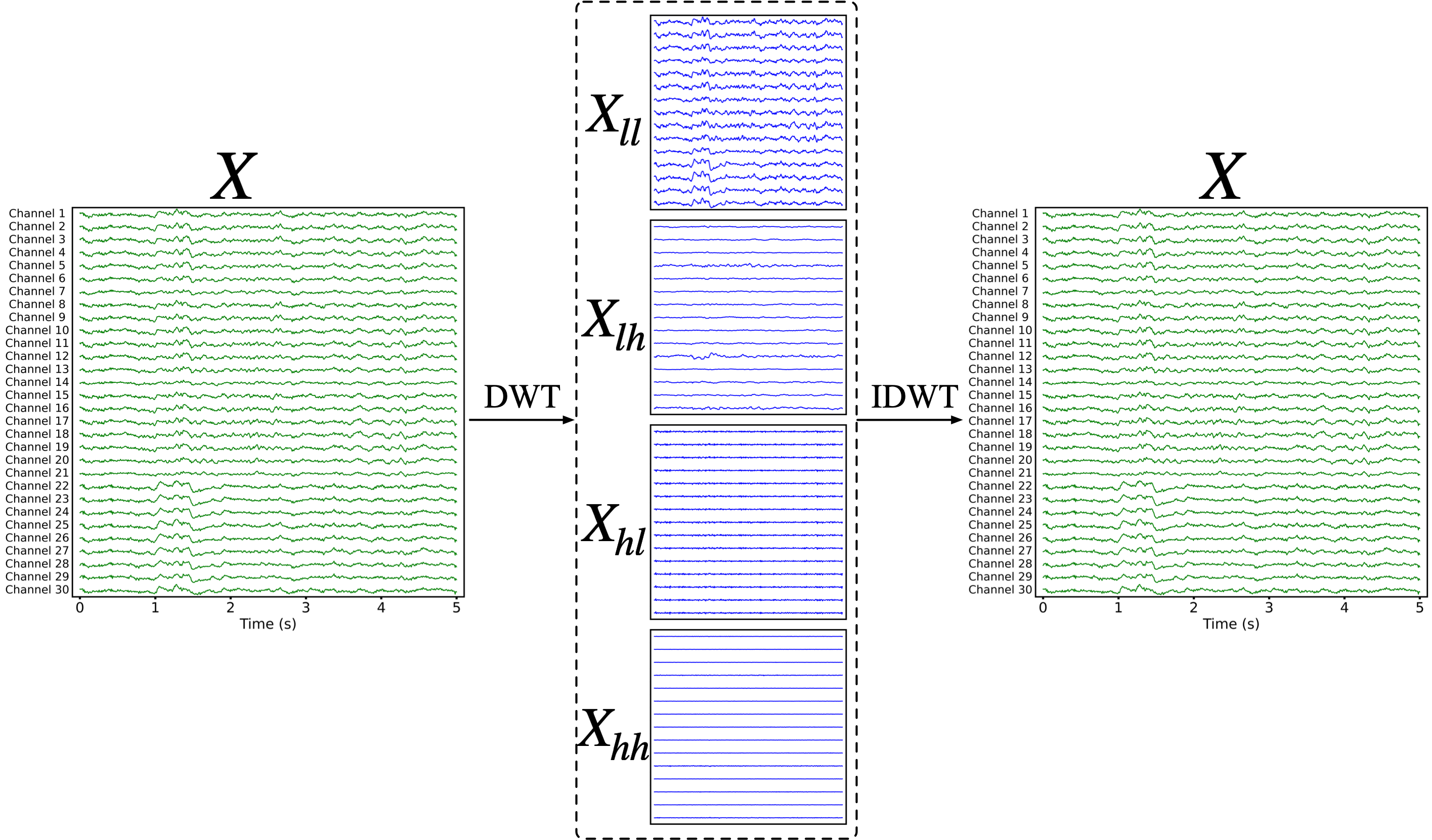}
  \captionsetup{font=small}
  \caption{Illustration of the EEG signal decomposition and reconstruction using DWT and IDWT.}
  \label{fig:dwt_detail}
\end{figure}

Due to the non-stationarity of EEG signals, wavelet transform is widely used in the time-frequency analysis of EEG signals. As is shown in Fig.~\ref{fig:dwt_detail}, wavelet transform enables the decomposition of the EEG signals into different frequency bands. To iteratively optimize the frequency-domain perturbations on EEG signals, the Discrete Wavelet Transform (DWT) is employed on the input data, decomposing the signal into a low-frequency component ($x_{ll}$) and three high-frequency components ($x_{lh}$, $x_{hl}$, $x_{hh}$) as:
\begin{equation}\label{eq2}
\begin{aligned}
    {x}_{l l}={L}{x}{L}^T&,{x}_{l h}={H}{x}{L}^T, \\
    {x}_{h l}={L}{x}{H}^T&,{x}_{h h}={H}{x}{H}^T,
\end{aligned}  
\end{equation}
where $L$ are the low-pass filters and $H$ are the high-pass filters of the orthogonal wavelet. Subsequently, the Inverse Discrete Wavelet Transform (IDWT) is utilized to reconstruct the original signal from the low-frequency and high-frequency components as:
\begin{equation}\label{3}
    {x}={L}^T{ x}_{l l}{L}+{H}^T{x}_{l h}{L}+{L}^T{x}_{h l}{H}+{H}^T{x}_{h h}{H}.
\end{equation}

\begin{figure*}[h!]
  \centering
  \includegraphics[width=\linewidth]{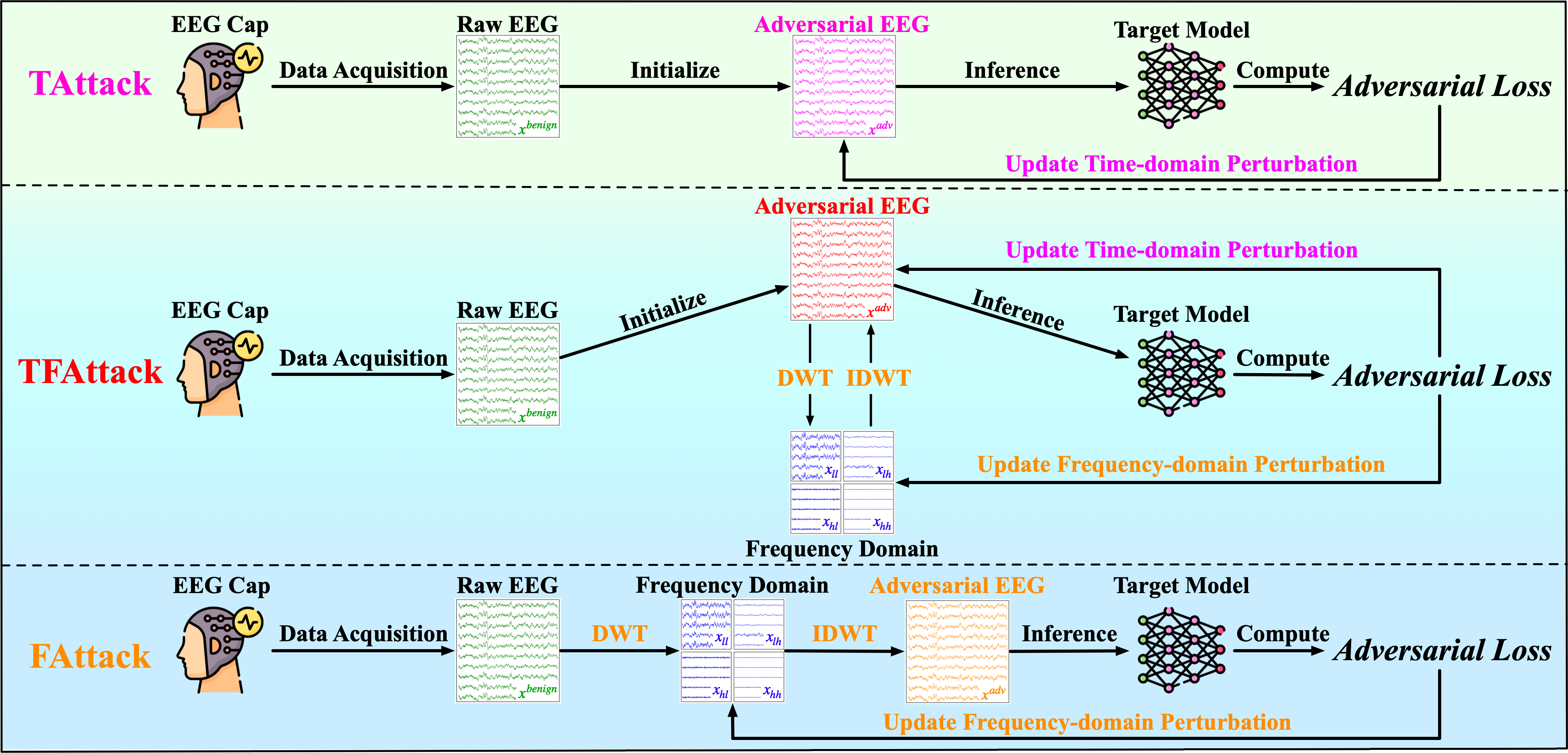}
  \captionsetup{font=small}
  \caption{
  Flowchart of the proposed TFAttack (middle) as a combination of TAttack (top) and FAttack (bottom).
  TAttack directly updates perturbations on raw EEG samples in the time domain, while FAttack first converts the samples to the frequency domain with DWT, updates the perturbation in the frequency domain and obtains the adversarial examples with IDWT.
  As for TFAttack, we have TAttack and FAttack take turns attacking the EEG samples, with TAttack first attacking the time-domain signal, and FAttack then attacking in the frequency domain. 
  }
\label{fig:tfattack}
\end{figure*}

Following the implementation of \cite{liWaveletIntegratedCNNs2020a}, both DWT and IDWT support gradient computation to facilitate backpropagation. For simplicity, we define
\begin{equation}\label{eq4}
\begin{aligned}
    &\mathrm{DWT}(x) = x_{ll}, x_{lh}, x_{hl}, x_{hh}, \\ 
    &\mathrm{IDWT}(x_{ll}, x_{lh}, x_{hl}, x_{hh}) = x.
\end{aligned}
\end{equation}

The pipeline of the FAttack is depicted in Fig.~\ref{fig:tfattack}. We summarize the optimization procedures as follows:
\begin{equation}\label{eq5}
\resizebox{.9\hsize}{!}{$
\begin{aligned}
    x^{\mathrm{adv}}_0&=x^{\mathrm{benign}},\\
    x^{\mathrm{adv}}_{t+1}&=\mathrm{IDWT}\Big(\mathrm{DWT}(x^{\mathrm{adv}}_{t}) \\
    &+{\alpha}{\cdot}{\nabla}_{\mathrm{DWT}(x^{\mathrm{adv}}_{t})}\mathcal{L }\left(f\left(\mathrm{IDWT}(\mathrm{DWT}(x^{\mathrm{adv}}_{t}))\right),y^{\mathrm{gt}}\right)\Big),
\end{aligned}
$}
\end{equation}
where $f$ represents the target model, $\mathcal{L}$ means the adversarial loss function and $\alpha$ denotes the step size for optimizing perturbation. We adopt part of the objective function of C\&W\cite{carliniEvaluatingRobustnessNeural2017} as our adversarial loss function.
\begin{equation}\label{eq6}
\resizebox{.9\hsize}{!}{$
\mathcal{L}\left(x, y^{gt}\right)=\max \left(Z\left(x\right)_{y^{gt}}-\max \left\{Z\left(x\right)_i: i \neq y^{gt}\right\},0\right),
$}
\end{equation}
where $Z(x)$ is logits, namely the output of the model except the softmax, and the $y^{gt}$ is the true label. The pseudocode of FAttack is presented in the supplementary materials.

\subsection{Time-Frequency Domain Attack}
To better harness both temporal and frequency information for crafting imperceptible effective adversarial examples, we devised TFAttack grounded in the time domain and frequency domain. 
By iteratively computing gradients on both time-domain and frequency-domain signals of the input data for perturbation updates, this approach combines the robust performance of time-domain attacks with the covert nature of frequency-domain attacks.
Furthermore, it demonstrates enhanced transferability of adversarial examples. The Fig.~\ref{fig:tfattack} illustrates the overall workflow of the time-frequency attack. The optimization procedures are summarized as follows and the pseudocode is presented in the supplementary materials:

\vspace{-1em}
\begin{equation}\label{eq7}
\resizebox{0.9\hsize}{!}{$
\begin{aligned}
    x^{\mathrm{adv}}_0&=x^{\mathrm{benign}},\\
    x^{\mathrm{adv}}_{t+1}&=x^{\mathrm{adv}}_{t}+{\alpha}{\cdot}{\nabla}_{x^{\mathrm{adv}}_{t}}\mathcal{L}\left(f\left(x^{\mathrm{adv}}_{t}\right),y^{\mathrm{gt}}\right), \\
    x^{\mathrm{adv}}_{t+2}&=\mathrm{IDWT}\Big(\mathrm{DWT}(x^{\mathrm{adv}}_{t+1}) \\
    &+{\alpha}{\cdot}{\nabla}_{\mathrm{DWT}(x^{\mathrm{adv}}_{t+1})}\mathcal{L}\left(f\left(\mathrm{IDWT}(\mathrm{DWT}(x^{\mathrm{adv}}_{t+1}))\right),y^{\mathrm{gt}}\right)\Big). \\
\end{aligned}
$}
\end{equation}

\section{Experiment}

\subsection{Experimental Setup}

\textbf{Datasets:}
We validated our method across three datasets: SEED Emotional Dataset (15 subjects)\cite{zhengEmotionMeterMultimodalFramework2019}, MTED Multitask EEG Dataset (30 subjects)\cite{kumarEvidenceTaskIndependentPersonSpecific2021}, and EEGMMI (109 subjects)\cite{goldbergerPhysioBankPhysioToolkitPhysioNet2000}. More information about the datasets can be found in supplementary materials.
For all datasets, We extracted 5 seconds of data from each sample as brainprint data and applied a bandpass filter of 1-79Hz for preprocessing.

\textbf{Evaluation Metrics:}
We primarily select
the Attack Success Rate (ASR) to measure the ratio of successfully attacked samples to the total samples. Regarding attack imperceptibility, we employ DTW score\cite{cuturiSoftDTWDifferentiableLoss2017}, $L_{2}$ norm, and cosine similarity to measure the visibility of adversarial examples to HVS. $L_{2}$ norm measures the pixel-level differences, but HVS is more sensitive to the pattern/shape of time-series data. Thus, we also adopt the DTW score, which better reflects HVS's preferences as one of the perceptual distances.

\textbf{Target Models:}
We selected three popular deep learning models specifically designed for EEG decoding as our target models: EEGNet\cite{lawhernEEGNetCompactConvolutional2018}, ShallowConvNet\cite{schirrmeisterDeepLearningConvolutional2017}, and DeepConvNet\cite{schirrmeisterDeepLearningConvolutional2017}. All three models were trained and tested on three datasets, with accuracy presented in the supplementary materials. 
\textbf{EEGNet}\cite{lawhernEEGNetCompactConvolutional2018}: EEGNet is a streamlined CNN model with a parameter count hovering around 1,000. This model comprises an initial input layer, two convolutional layers, and a classification layer. It is worth noting that EEGNet adopts depthwise separable convolution to replace the conventional convolution.
\textbf{DeepConvNet}\cite{schirrmeisterDeepLearningConvolutional2017}: DeepConvNet has four convolutional layers and a classification layer with more parameters. Its first convolutional layer is specifically designed for EEG data, while subsequent layers follow a more general format.
\textbf{ShallowConvNet}\cite{schirrmeisterDeepLearningConvolutional2017}: Stemming from the DeepConvNet and influenced by the filter bank common spatial pattern\cite{angFilterBankCommon2008}, ShallowConvNet is a more shallow CNN. The convolution layers are similar to DeepConvNet's first layer but differ in kernel size, activation functions, and pooling layers.

\begin{table*}[h!]
\caption{White-box attack performance comparison on three datasets with three target models.}
\label{table3}
\centering

\begin{tabular*}{\linewidth}{@{\extracolsep{\fill}}cccccccc}
\toprule
Dataset                  & Model                           & Metrics            & FGSM\cite{goodfellowExplainingHarnessingAdversarial2015a}      & BIM\cite{kurakinAdversarialExamplesPhysical2017a}       & PGD\cite{madryDeepLearningModels2018}       & C\&W\cite{carliniEvaluatingRobustnessNeural2017}       & TFAttack(ours) \\ \midrule
\multirow{12}{*}[-2ex]{EEGMMI} & \multirow{4}{*}{DeepConvNet}    & ASR(\%) $\uparrow$  & 67.23     & 81.30     & 82.14     & 76.47      & \textbf{97.27}          \\ 
                         &                                 & $L_2$$ \downarrow$     & 377.82    & 354.90    & 363.13    & 14523.98   & \textbf{263.34}         \\
                         &                                 & DTW$ \downarrow$    & 83215.74  & 76424.18  & 78150.81  & 1796593.32 & \textbf{51388.50}       \\
                         &                                 & Cosine Similarity$ \uparrow$ & 0.99846   & 0.99869   & 0.99863   & 0.52394    & \textbf{0.99886}        \\ \cmidrule{2-8}
                         & \multirow{4}{*}{EEGNet}         & ASR(\%) $\uparrow$  & 87.79     & 93.68     & 94.11     & 81.26      & \textbf{99.37}          \\
                         &                                 & $L_2$ $\downarrow$     & 381.96    & 363.49    & 381.23    & 13402.12   & \textbf{208.92}         \\
                         &                                 & DTW $\downarrow$    & 86103.23  & 81257.84  & 85166.86  & 1640992.31 & \textbf{42506.25}       \\
                         &                                 & Cosine Similarity$ \uparrow$ & 0.99856   & 0.99874   & 0.99862   & 0.53063    & \textbf{0.99930}        \\ \cmidrule{2-8}
                         & \multirow{4}{*}{ShallowConvNet} & ASR(\%) $\uparrow$  & 81.22     & 89.76     & 90.49     & 48.78      & \textbf{96.34}          \\
                         &                                 & $L_2$ $\downarrow$     & 375.87    & 331.16    & 337.82    & 14858.79   & \textbf{204.24}         \\
                         &                                 & DTW $\downarrow$    & 84178.24  & 71570.65  & 72950.91  & 1833295.95 & \textbf{40956.35}       \\
                         &                                 & Cosine Similarity $\uparrow$ & 0.99861   & 0.99889   & 0.99884   & 0.49172    & \textbf{0.99924}        \\ \midrule
\multirow{12}{*}[-2ex]{SEED}   & \multirow{4}{*}{DeepConvNet}    & ASR(\%) $\uparrow$  & 66.13     & 78.16     & 78.76     & 37.88      & \textbf{95.99}          \\
                         &                                 & $L_2$ $\downarrow$     & 338.41    & 305.52    & 315.36    & 30695.48   & \textbf{208.50}         \\
                         &                                 & DTW $\downarrow$     & 74723.62  & 66285.13  & 68317.49  & 7227219.34 & \textbf{43119.95}       \\
                         &                                 & Cosine Similarity $\uparrow$ & 0.98714   & 0.98909   & 0.98865   & 0.03197    & \textbf{0.99040}        \\ \cmidrule{2-8}
                         & \multirow{4}{*}{EEGNet}         & ASR(\%) $\uparrow$  & 56.22     & 70.08     & 70.48     & 61.65      & \textbf{96.79}          \\
                         &                                 & $L_2$ $\downarrow$     & 315.58    & 295.75    & 299.74    & 32953.53   & \textbf{182.17}         \\
                         &                                 & DTW $\downarrow$    & 74504.88  & 68961.63  & 69845.78  & 7845750.83 & \textbf{39026.99}       \\
                         &                                 & Cosine Similarity $\uparrow$ & 0.98829   & 0.98859   & 0.98832   & 0.05030    & \textbf{0.99135}        \\ \cmidrule{2-8}
                         & \multirow{4}{*}{ShallowConvNet} & ASR(\%) $\uparrow$  & 70.42     & 91.46     & 92.08     & 62.92      & \textbf{94.17}          \\
                         &                                 & $L_2$ $\downarrow$     & 321.89    & 280.48    & 284.58    & 19377.20   & \textbf{196.78}         \\
                         &                                 & DTW $\downarrow$    & 76025.90  & 64774.03  & 65668.36  & 4257921.83 & \textbf{42235.76}       \\
                         &                                 & Cosine Similarity $\uparrow$ & 0.98647   & 0.98833   & 0.98811   & 0.21241    & \textbf{0.99179}        \\ \midrule
\multirow{12}{*}[-2ex]{MTED}   & \multirow{4}{*}{DeepConvNet}    & ASR(\%) $\uparrow$  & 75.88     & 82.83     & 82.83     & 81.82      & \textbf{86.36}          \\
                         &                                 & $L_2$ $\downarrow$     & 315.06    & 198.50    & 198.49    & 4280.06    & \textbf{151.09}         \\
                         &                                 & DTW $\downarrow$    & 117562.12 & 72989.85  & 72987.81  & 1097841.39 & \textbf{50996.75}       \\
                         &                                 & Cosine Similarity $\uparrow$ & 0.98456   & 0.99376   & 0.99376   & 0.50125    & \textbf{0.99428}        \\ \cmidrule{2-8}
                         & \multirow{4}{*}{EEGNet}         & ASR(\%) $\uparrow$  & 73.08     & 85.16     & 85.71     & 86.81      & \textbf{87.36}          \\
                         &                                 & $L_2$ $\downarrow$     & 204.15    & 182.11    & 184.84    & 4844.58    & \textbf{139.70}         \\
                         &                                 & DTW $\downarrow$    & 78960.68  & 70214.47  & 71222.30  & 1355980.84 & \textbf{49014.23}       \\
                         &                                 & Cosine Similarity $\uparrow$ & 0.99282   & 0.99395   & 0.99379   & 0.46466    & \textbf{0.99463}        \\ \cmidrule{2-8}
                         & \multirow{4}{*}{ShallowConvNet} & ASR(\%) $\uparrow$  & 75.88     & 85.93     & 84.92     & 51.76      & \textbf{92.96}          \\
                         &                                 & $L_2$ $\downarrow$     & 291.43    & 292.48    & 290.12    & 4488.82    & \textbf{236.69}         \\
                         &                                 & DTW $\downarrow$    & 109559.61 & 106878.82 & 106086.97 & 1181837.78 & \textbf{80188.48}       \\
                         &                                 & Cosine Similarity $\uparrow$ & 0.98666   & 0.98603   & 0.98621   & 0.47316    & \textbf{0.98734}      
                         \\ \bottomrule
\end{tabular*}
\end{table*}

\textbf{Implementation details:}
We selected the Adam~\cite{kingmaAdamMethodStochastic2015} optimizer for TFAttack, setting the learning rate to 0.02. The Haar wavelet~\cite{stankovicHaarWaveletTransform2003a} was chosen as the wavelet basis function for DWT and IDWT. Implementation details of comparative methods are presented in the supplementary materials. We randomly selected 500 samples on each test set.

\subsection{Experiments and Results}
We conducted four experiments, namely white-box attack, perception study, transferability, and ablation study. Due to the page limit, we present the ablation study and its results in the supplementary materials.

\textbf{White-box Attack:} 
In this experiment, we verified the attack strength and imperceptibility of the adversarial examples generated by TFAttack. White-box attacks were carried out on EEGMMI\cite{goldbergerPhysioBankPhysioToolkitPhysioNet2000}, SEED\cite{zhengEmotionMeterMultimodalFramework2019} and MTED datasets\cite{kumarEvidenceTaskIndependentPersonSpecific2021}, and compared TFAttack with methods such as FGSM\cite{goodfellowExplainingHarnessingAdversarial2015a}, BIM\cite{kurakinAdversarialExamplesPhysical2017a}, PGD\cite{madryDeepLearningModels2018}, and C\&W\cite{carliniEvaluatingRobustnessNeural2017}. Table~\ref{table3} shows that TFAttack has the lowest $L_{2}$ and DTW scores and the highest cosine similarity compared with other attack methods, and has an attack effect on all three datasets. At the same time, TFAttack has the highest ASR in SEED ,MTED, and EEGMMI dataset, which indicates that TFAttack is more effective than other attack methods.

\begin{figure*}[h!]
    \begin{minipage}{0.49\linewidth}
        \centering
        \includegraphics[width=\linewidth]{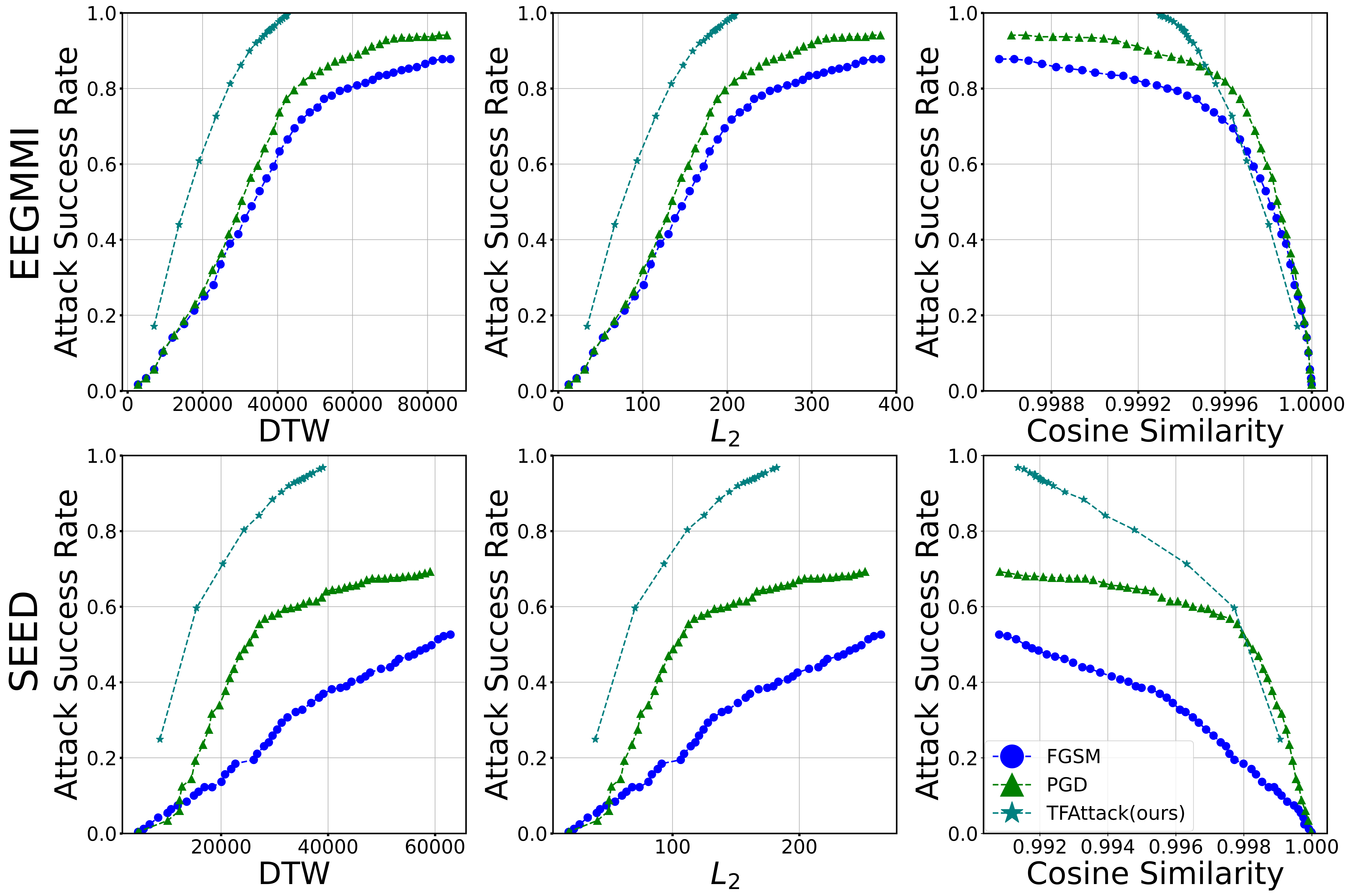}
        \caption{Perception study carried out on EEGMMI and SEED with EEGNet as the target model.}
        \label{fig:perception study}
    \end{minipage}
    \hfill
    \begin{minipage}{0.49\linewidth}
        \makeatletter\def\@captype{table}\makeatother %
        \caption{The attack success rates (\%) $\uparrow$ / DTW $\downarrow$ of transferring adversarial examples across three models in EEGMMI.}
        \label{table4}
        \centering
        \scalebox{0.65}{
        \renewcommand\arraystretch{1.5}
        \begin{tabular}{@{\extracolsep{\fill}}ccccc}
        \toprule
        \multirow{2}{*}{Substitute Model} & \multirow{2}{*}{Attack} & \multicolumn{3}{c}{Target Model}                                          \\ \cline{3-5} 
                                                             &                         & DeepConvNet            & EEGNet                 & ShallowConvNet          \\ \midrule
        \multirow{3}{*}{DeepConvNet}                         & FGSM\cite{goodfellowExplainingHarnessingAdversarial2015a}                    & -/-                    & 3.78/69449.73          & 20.80/64060.01          \\
                                                             & PGD\cite{madryDeepLearningModels2018}                     & -/-                    & 3.57/59221.22          & 21.01/58274.48          \\
                                                             & TFAttack(ours)          & -/-                    & \textbf{3.78/41070.07} & \textbf{22.48/49129.50} \\ \midrule
        \multirow{3}{*}{EEGNet}                              & FGSM\cite{goodfellowExplainingHarnessingAdversarial2015a}                    & 4.00/53269.37          & -/-                    & \textbf{19.79}/52454.03          \\
                                                             & PGD\cite{madryDeepLearningModels2018}                     & 4.00/52292.83          & -/-                    & 19.58/53947.74          \\
                                                             & TFAttack(ours)          & \textbf{4.00/27737.71} & -/-                    & 19.16/\textbf{37609.36} \\ \midrule
        \multirow{3}{*}{ShallowConvNet}                      & FGSM\cite{goodfellowExplainingHarnessingAdversarial2015a}                    & 4.63/83430.45          & \textbf{5.61}/96061.25          & -/-                     \\
                                                             & PGD\cite{madryDeepLearningModels2018}                     & 4.88/76570.38          & 5.37/78922.13          & -/-                     \\
                                                             & TFAttack(ours)          & \textbf{5.12/29061.51} & 4.63/\textbf{38273.96} & -/-                     \\ \bottomrule
        \end{tabular}
        }
    \end{minipage}
\end{figure*}

\textbf{Perception Study:} 
To further investigate the rate at which the ASR increases with the enlargement of perceptual distances, we adjusted the perturbation thresholds $\epsilon$ for FGSM\cite{goodfellowExplainingHarnessingAdversarial2015a} and PGD\cite{madryDeepLearningModels2018}, as well as the iteration number for TFAttack, to increase the perceptual distances of perturbations, and record the corresponding ASR. We tested the attack performance on EEGNet using two datasets, employing $L_{2}$, DTW, and cosine similarity as measures of perceptual distances. Lower $L_{2}$ and DTW values imply smaller perturbations, and higher cosine similarity indicates a closer match between original and adversarial samples. Fig.~\ref{fig:perception study} demonstrates that ASR rises with the strength of perturbations. Additionally, TFAttack consistently generates more powerful adversarial examples at the same perceptual distances or achieves the same ASR with lower $L_{2}$, DTW, and higher cosine similarity. This verified TFAttack's capability to generate both powerful and invisible adversarial examples.

\textbf{Transferability:}
To study the transferability of adversarial examples generated by the proposed TFAttack, we also conducted grey-box attack experiments on the EEGMMI dataset. In the grey-box attack scenario, the attacker has the dataset of the target model and uses the alternative model to generate adversarial examples. Therefore, we first used the training set of EEGMMI to train three substitute models, then used FGSM\cite{goodfellowExplainingHarnessingAdversarial2015a}, PGD\cite{madryDeepLearningModels2018} and TFAttack to generate adversarial examples on the test set, and then attacked the three target models respectively to obtain ASR and DTW scores. According to Table~\ref{table4}, we can find that the robustness of ShallowConvNet is poor, and the DTW score of TFAttack is much lower than that of FGSM and PGD when ASR is not much different from FGSM and PGD. Therefore, the transferability of TFAttack is better than that of FGSM and PGD.

\section{Conclusion}
We propose an imperceptible adversarial attack method in both time and frequency domains, which successfully addresses the challenges of insufficient frequency information utilization and the visibility of adversarial examples inherent in existing methods. We achieved an efficient transition between EEG's time and frequency domains by utilizing wavelet transformation, thereby generating more imperceptible adversarial examples through alternating optimization. Extensive experimental results validate the superior attack performance of our algorithm across various datasets and models. In summary, this study reveals the vulnerabilities of deep-learning models in brainprint recognition systems. We hope our work will draw more attention to the robustness research in brainprint recognition.

\bibliographystyle{IEEEtran}
\bibliography{final}

\newpage
\renewcommand{\appendixname}{Supplementary Material}
\appendix


\let\OLDthebibliography\thebibliography
\renewcommand\thebibliography[1]{
  \OLDthebibliography{#1}
  \setlength{\parskip}{0pt}
  \setlength{\itemsep}{0pt plus 0.3ex}
}

\def\x{{\mathbf x}}
\def\L{{\cal L}}

\vspace{1em}
\textbf{\normalsize 1. Algorithm Pseudocode}
\vspace{1em}

To better illustrate the processes of our proposed FAttack and TFAttack methods, here are the pseudocodes for FAttack and TFAttack. All the equations below can be found in the main text.

\renewcommand{\algorithmicrequire}{\textbf{Input:}}
\renewcommand{\algorithmicensure}{\textbf{Output:}}
\begin{algorithm}[H]
\caption{FAttack.}\label{alg:alg1}
\begin{algorithmic}[1]
\REQUIRE A benign EEG sample: $x^{benign}$, corresponding ground truth label:$y^{gt}$ and a target model $f$;
\ENSURE A adversarial EEG sample: $x^{adv}$;
\STATE  Initialization: $x_0^{adv} = x^{benign}$\;
\FOR{$n=0$ to $N$}
    \STATE Calculate the frequency domain data $DWT(x_{n}^{adv})$ as Eq. 4;
    \STATE Apply the IDWT on frequency domain data to get the time domain data $x_{n}^{adv}$ as Eq. 4 shows;
    \STATE Inference on time domain data $x_{n}^{adv}$ using target model $f$ and calculate adversarial loss $\mathcal{L}$; 
    \STATE Update the adversarial example $x_{n+1}^{adv}$ by Eq. 5;
\ENDFOR

\STATE \textbf{return}  $x_{N}^{adv}$
\end{algorithmic}
\label{alg1}
\end{algorithm}

\renewcommand{\algorithmicrequire}{\textbf{Input:}}
\renewcommand{\algorithmicensure}{\textbf{Output:}}
\begin{algorithm}[H]
\caption{TFAttack.}\label{alg:alg2}
\begin{algorithmic}[1]
\REQUIRE A benign EEG sample: $x^{benign}$, corresponding ground truth label:$y^{gt}$ and a target model $f$;
\ENSURE An adversarial EEG sample: $x^{adv}$;
\STATE  Initialization: $x_0^{adv} = x^{benign}$\;
\FOR{$n=0$ to $N$}
    \STATE Inference on time domain data $x_{n}^{adv}$ using target model $f$ and calculate adversarial loss $\mathcal{L}$; 
    \STATE Update the adversarial example $x_{n+1}^{adv}$ in time domain by Eq. 7;
    \STATE Calculate the frequency domain data $DWT(x_{n}^{adv})$ as Eq. 4 shows;
    \STATE Apply the IDWT on frequency domain data to get the time domain data $x_{n+1}^{adv}$ as Eq. 4 shows;
    \STATE Inference on time domain data $x_{n+1}^{adv}$ using target model $f$ and calculate adversarial loss $\mathcal{L}$; 
    \STATE Update the adversarial example $x_{n+2}^{adv}$ by Eq. 7;
\ENDFOR

\STATE \textbf{return}  $x_{N}^{adv}$
\end{algorithmic}
\label{alg2}
\end{algorithm}

\vspace{1em}
\textbf{\normalsize 2. Datasets}
\vspace{1em}

We conducted experiments on three datasets: EEGMMI\cite{goldbergerPhysioBankPhysioToolkitPhysioNet2000}, SEED\cite{zhengEmotionMeterMultimodalFramework2019}, and MTED\cite{kumarEvidenceTaskIndependentPersonSpecific2021}. The basic information in presented in Table ~\ref{tab1}.

\textbf{EEGMMI\cite{goldbergerPhysioBankPhysioToolkitPhysioNet2000}:} This dataset consists of motor imagery data from 109 subjects, including rest state, hand-foot motor imagery, and left-right hand motor imagery. Each subject completed 14 sessions with 180 samples each. The EEG data were collected using a 64-channel system following the 10-10 system, with a sampling rate of 160Hz.

\textbf{SEED\cite{zhengEmotionMeterMultimodalFramework2019}:} This dataset contains emotional elicitation data from 15 subjects, including four emotional states (i.e., Happy, Sad, Neutral, Fear). Each subject participated in experiments over three sessions on different days, with data captured by a 62-channel EEG cap at a sampling rate of 200Hz.

\textbf{MTED\cite{kumarEvidenceTaskIndependentPersonSpecific2021}:} The dataset includes multi-task data from 30 subjects, including 12 tasks. Subjects contributed data from at least two to a maximum of five days, with each session containing up to four tasks. The data were recorded using a high-density 128-channel EEG device, at a sampling rate of 250Hz.

\begin{table}[t]
\caption{Datasets information.}
\label{tab1}
\centering
\scalebox{0.85}{
\begin{tabular}{@{\extracolsep{\fill}}ccccc}
\toprule
Datasets & subjects & channels & sampling rate & paradigm    \\ \midrule
EEGMMI\cite{goldbergerPhysioBankPhysioToolkitPhysioNet2000}   & 109      & 64       & 160           & Motor Imagery          \\
SEED\cite{zhengEmotionMeterMultimodalFramework2019}     & 15       & 62       & 200           & Emotion     \\
MTED\cite{kumarEvidenceTaskIndependentPersonSpecific2021}     & 30       & 128      & 250           & Multiple Tasks \\
\bottomrule
\end{tabular}
}
\end{table}

\vspace{1em}
\textbf{\normalsize 3. Target Models}
\vspace{1em}

As DeepConvNet\cite{schirrmeisterDeepLearningConvolutional2017}, EEGNet\cite{lawhernEEGNetCompactConvolutional2018}, and ShallowConvNet\cite{schirrmeisterDeepLearningConvolutional2017} are frequently used as the backbone for feature extraction in various deep learning models, we select them to be the target models. And we trained and tested them on EEGMMI, SEED, and MTED datasets. Every combination of models and datasets achieved a high brainprint recognition accuracy, as shown in Table ~\ref{tab2}.
\begin{table}[htbp]
\caption{Accuracy of DeepConNet, EEGNet, and ShallowConvNet on EEGMMI, SEED, and MTED datasets.}
\label{tab2}
\centering
\begin{tabular}{@{\extracolsep{\fill}}cccc}
\toprule
Model          & EEGMMI & SEED   & MTED   \\ \midrule
DeepConvNet\cite{schirrmeisterDeepLearningConvolutional2017}    & 0.9607 & 0.9951 & 0.9900 \\
EEGNet\cite{lawhernEEGNetCompactConvolutional2018}         & 0.9535 & 0.9941 & 0.9100 \\
ShallowConvNet\cite{schirrmeisterDeepLearningConvolutional2017} & 0.8150 & 0.9704 & 0.9950 \\
\bottomrule
\end{tabular}
\end{table}

\vspace{1em}
\textbf{\normalsize 4. Implementation Details of Other Methods}
\vspace{1em}

We selected FGSM\cite{goodfellowExplainingHarnessingAdversarial2015a}, BIM\cite{kurakinAdversarialExamplesPhysical2017a}, PGD\cite{madryDeepLearningModels2018}, and C\&W\cite{carliniEvaluatingRobustnessNeural2017} as comparative attack methods. FGSM, BIM, and PGD utilize a perturbation budget ($\epsilon$) to control the strength of the perturbation. Due to the non-stationarity of EEG signals, we set the perturbation budget ($\epsilon$) for each EEG sample as the product of the perturbation budget ($\epsilon$) and the sample's variance ($\sigma^2$). 
\begin{equation}
\epsilon^{\prime}=\epsilon\cdot\sigma^{2}
\end{equation}
The step sizes for BIM and PGD are set to one-tenth of the perturbation budget. 
\begin{equation}
stepsize=0.1\cdot\epsilon^{\prime}
\end{equation}
Similarly to TFAttack, C\&W also employs the Adam optimizer, with the learning rate set at 0.002, the parameter $c$ at 1, and $\kappa$ at 0. We performed max-min normalization on the EEG samples to prevent floating-point overflow issues associated with the tanh (hyperbolic tangent function) in the C\&W method.

\vspace{1em}
\textbf{\normalsize 5. Ablation Study}
\vspace{1em}

In order to study the contribution of TAttack and FAttack in TFAttack, we use them to attack EEGNet on EEGMMI dataset. Considering that TFAttack involves two separate attacks on the time and frequency domains in one iteration, we set the learning rates for TAttack and FAttack at 0.02, and for TFAttack at 0.01. As observed in Table~\ref{table3}, TAttack exhibits a higher ASR but with larger perceptual distances of perturbations, whereas FAttack shows the opposite trend. TFAttack effectively leverages the advantages of both TAttack and FAttack, achieving the highest ASR with a relatively smaller perceptual distance. The DTW score of TFAttack is higher than that of FAttack, which may be due to the effect of time-domain signal being attacked in TFAttack.

\begin{table}[h!]
\caption{The ablation study carried out on EEGMMI with EEGNet as the target model.}
\centering
\label{table3}
\scalebox{0.9}{
\begin{tabular}{@{\extracolsep{\fill}}ccccc}
\toprule
Attack   & ASR(\%) & $L_s$     & DTW       & Consine Similarity \\ \midrule
TAttack  & 94.11   & 158.18 & 35099.39  & 0.99957             \\
FAttack  & 93.68   & 161.43 & \textbf{28011.96}  & 0.99955             \\
TFAttack & \textbf{94.53}   & \textbf{147.26} & 29999.00  & \textbf{0.99963}             \\ 
\bottomrule
\end{tabular}
}
\end{table}

\end{document}